# Hi Gemini, Who Should I Vote For? The Benefits (and Problems) of Using LLMs As Voting Advice Applications

by

William John Atkinson

Department of Politics, LUISS University

and

Paul Saproza

Department of Politics, LUISS University

18 November, 2026

**Contact information**

William John Atkinson (corresponding author)
Viale Romania, 32
00197
Roma RM
Italy
watkinson@luiss.it

Paul Saproza
Viale Romania, 32
00197
Roma RM
Italy

**Hi Gemini, Who Should I Vote For? The Benefits (and Problems) of Using LLMs As Voting Advice Applications**

**Abstract**

Large language models (LLMs) can provide voters information about political issues, party positions, and even advise on which party to vote for. However, no previous study explores whether off-the-shelf LLMs could be used as voting advice applications (VAAs) to improve voter-party congruence. In this article, we address this gap by examining how Gemini 3 Flash advises a sample of 6,022 German survey respondents on which party to vote for, based on their policy preferences and priorities alone. In brief, we find that for most respondents, following the LLM's advice would benefit voter-party policy alignment (i.e. congruence), with a significantly stronger remedial effect among voters with low levels of political interest and formal education. However, our analysis also reveals concerning biases in the LLM's recommendations. For instance, the LLM suggested voting for the AfD at ~1.5 times the frequency of a standard salience-weighted proximity model. In addition, the LLM hardly ever advised voting for either Die Linke or the BSW. We conclude that while LLMs can approximate the behaviour of VAAs, users should be aware that the LLM's recommendations are still biased in ways that are extremely difficult to trace.

## Introduction

Citizens have a need for clear guidance at election time. Previous work shows that voters are often poorly informed about political issues, and where parties stand on them (Friedman, 2006; Hardin, 2009; Mulligan, 2017). Becoming politically informed is considerably more costly than voting, with an uncertain payoff, such that mass democracies have been characterized as "ignorant democracies" (Hardin, 2009). Voters in large societies face rational ignorance because learning about complex political issues takes too much time and effort, and a single vote almost never changes an election, meaning individuals have no selfish reason to gather this information (Christiano, 2018; Downs, 1957).

As a result, voters can find it difficult to identify which party best represents their policy preferences, thereby reducing the quality of democratic outcomes. More specifically, previous studies show that *congruence* is poorer among less politically informed voters (Boonen et al., 2017; Carroll & Kubo, 2018; Enyedi et al., 2020; Lesschaeve, 2017; Rodríguez-Zepeda & Otero-Felipe, 2021), where 'congruence' is defined as the degree to which the voter agrees with the positions of the party they support (i.e. voter-party congruence; Dalton, 1985). Congruence is widely viewed as an indicator of democratic health – if voters do not agree with their party on key policy issues, then substantive representation suffers (Pitkin, 1967). In theory, voters who lack political information are more poorly equipped to choose a congruent party. In this regard, the decline of cleavage voting (Franklin, 1992) and partisan dealignment (Dalton et al., 2000) have probably made the situation worse, weakening the value of group-based heuristics (Garzia & Marschall, 2012; Scarrow, 2010).

This article considers whether publicly available LLMs, like Google's Gemini, can improve voter-party congruence, by advising their users which party to vote for in elections. In this sense, we expect large language models (LLMs) to function much like voting advice applications (VAAs), making a judgement about which party is most congruent with the voter's policy preferences. To begin, we argue that LLMs enjoy key advantages over VAAs – principally, that they are more likely to be accessed by precisely those voters who lack political information to start with. However, a major drawback is the lack of transparency. Whereas VAAs routinely make their methodology easily accessible, we cannot as easily (or, at all) explain the outputs of AI models (Rudin, 2019). As a result, it is important to examine empirically the recommendations made by LLMs in this context. Only by doing this can we make a judgment about whether LLMs would have net positive effect on voter decision-making, and therefore, voter-party congruence.

To achieve this, we propose a novel research design where the LLM (in this case, Gemini 3 Flash) is fed the policy preferences and priorities of 6,022 German voters who responded to the post-election edition of the German Longitudinal Election Study 2025. Based on this information, the LLM is required to recommend a party to each respondent. The LLM's recommendations are then compared side-by-side to a) the party the respondent actually voted for, and b) a recommendation designed to mirror a conventional VAA design – that is, a salience-weighted proximity model where party positions are taken from expert surveys ('Expert model'). Overall, we find that for a slim majority of respondents, the LLM recommends a more congruent party than the one actually supported. In only just over 10% of cases did LLM's recommendation put congruence at risk, and for just under 70% of respondents, the LLM's recommendation matched the Expert model. However, the frequency of the LLM's recommendations appears to be biased towards the AfD, and against smaller parties like Die Linke and the BSW. We conclude that while off-the-shelf LLMs can roughly mimic the function of a VAA, these apparent biases compel further investigation, and should be cause for concern for LLM developers, users, and political scientists.

## Using Large Language Models as Voting Advice Applications

Many citizens are looking for assistance to identify who they should vote for. VAAs have played this role at least since the middle of the 2000s (Breuer, 2010; Farrell & Schmitt-Beck, 2008;

Marzuca et al., 2011; Trechsel & Mair, 2011), facilitated by the emergence of the internet (Garzia & Marschall, 2012; Hooghe & Teepe, 2007). Following the development of machine learning through the 2010s and 2020s, culminating in the public release of ChatGPT in November 2022, a new tool is substantially transforming users' habits: large language models (LLMs).

Although VAAs and LLMs are very different tools, their use can overlap. On the one hand, VAAs are digital tools that match user's policy choices with the positions of political parties to suggest a vote (Gemenis, 2024). On the other hand, LLMs are computer programs that generate text by predicting the most probable continuation of a given input, on the basis of patterns learned from vast corpora during training (Von Der Heyde et al., 2025). As such, LLMs encompass a considerably wider range of potential applications than VAAs. Crucially however, LLMs could perform (or at least, imitate) the function of traditional VAAs, by providing citizens with assistance in identifying which party best aligns with their preferences (Garzia & Marschall, 2012). If they perform this function well, the widespread use of LLMs as VAAs might lead to substantially better voter-party congruence. However, we still do not know how well LLMs perform this function – or put differently, how well an LLM can imitate a VAA.

Moreover, some developers have constrained LLMs so as not to take clear political stances, including giving voting recommendations. For instance, in the run-up to the 2024 US presidential election, OpenAI published a blog post claiming that ChatGPT had been instructed to "not express political preferences or recommend candidates even when asked explicitly" (OpenAI, 2024). Nonetheless, when pressed directly, the LLMs we tested were usually willing to give such a recommendation. Therefore, with regards to political content, it appears that most developers have opted for "soft censorship" (Noels et al., 2025), whereby certain LLM behaviours are discouraged rather than being refused entirely. Although the scope and effectiveness of this censorship may vary across models and providers (Ahmed et al., 2025), the decision by LLM developers to censor political content is widespread, and a matter of public record (Munich Security Conference 2024). To summarise, developers probably perceive the use of LLMs for such purposes as controversial. Therefore, our inquiry is also relevant to the question of whether developers *should* censor the output of LLMs in this context.

*Advantages*

Indeed, upon reviewing the VAA literature, we identify several potential benefits of using LLMs for voting advice. First and foremost, by mimicking the function of a traditional VAA, the LLM circumvents the need for voters to learn about party positions, or party politics more broadly. In theory, it would be sufficient for a voter to provide the LLM with their policy preferences and priorities, and the predictive algorithm would tell the user which party, or parties, best serve their interests. This way, LLMs (and VAAs more broadly) improve the efficiency of the democratic process, ensuring that voters are matched with the party that best aligns with their own beliefs, and thereby improving voter-party congruence (Garzia & Marschall, 2012). In other words, these tools are designed to offset the negative effects of being politically uninformed.

Second, LLMs have a crucial advantage over VAAs, in that they are more likely to be accessed by the voters who need them most. A central benefit of VAAs is that they may facilitate participation, which is essential for democracy (Christiano, 2018; Dahl, 2020; Jacob, 2015; Mill, 1974; Pateman, 1975). In this respect, LLMs promise significant advances over traditional VAAs because they reach a considerably larger audience than VAAs currently do. VAAs only reach a narrow group of user (Garzia & da Silva, 2026; Garzia & Marschall, 2012). Instead, because they are embedded within everyday apps that millions already use—either dedicated apps such as Chatgpt, Claude, Grok, etc., but also services integrating it such as Google, Bing, Whatsapp, and so on (Cui &

Yasseri, 2024; Marcos & Juan C, 2024)—, LLMs can reach a much wider audience, especially in democratic countries (Liang et al., 2025). In addition, LLMs not only reach a greater number of people than VAAs, but are likely to include a more diverse audience. Indeed, VAAs are often criticized as being too elitist (Gemenis 2024). The most recent studies about VAA adoption confirm that there is little change in the sociodemographic composition of the population of VAA users despite continuous expansion (Garzia & da Silva, 2026). The people who used VAAs the most are young, highly educated, and very interested in politics, suggesting these tools predominately reach those who need the least help (Van de Pol et al., 2014). Because users can interact with an LLM in natural language, as they would with another person, LLMs are easier to use than the traditional VAAs, which force users to click through rigid and complicated surveys. Citizens of all backgrounds already use these tools for everyday choices, from seeking health advice to obtaining financial tips (Chen et al., 2024; Choudhury & Shamszare, 2023; Fletcher & Nielsen, 2024; Kasneci et al., 2023). As practical guidance is the most common use case of ChatGPT (Chatterji et al., 2025), users could easily turn to LLMs for electoral advice too.

Third, LLMs are responsive users' inputs in ways that traditional VAAs have not been. In fact, there have recently been efforts to integrate LLMs into VAAs, with chatbots assisting the user by rephrasing ambiguous questions, provide background details, and engage in iterative dialogue with the user until the user understands the issue (Kamoen & Liebrecht, 2022; van Zanten & Boumans, 2023). Moreover, LLMs could be used to make VAAs themselves more democratic (i.e. user-driven). An enduring problem of traditional VAAs is that their creators must impose a battery questions on the respondent, thereby intervening in the voter's decision by deciding *a priori* which issues are salient (Gemenis, 2024). By contrast, the LLM as VAA approach can afford to be much more open-ended, with the user stating from the outset which issues matter most to them personally (Fish et al., 2023).

*Disadvantages*

However, the use of LLMs as VAAs comes with several downsides. The most urgent of these relates to the transparency of the LLM's matching algorithm – that is, the procedure by which the recommended party is determined for a given respondent. With VAAs, the methodology used to match respondents with parties is nearly always made explicit. For instance, the German *Wahl-O-Mat*, one of the most widely used VAAs, uses a sort of salience-weighted proximity model (HHU Düsseldorf, 2026). In more general terms, the choice of algorithm is sure to affect the recommendations made by the VAA (Louwerse & Rosema, 2014; Wagner & Ruusuvirta, 2012). Like VAAs, LLMs do tend to 'reason' in terms of voters' policy preferences and priorities, performing something akin to a spatial matching algorithm. However, because LLMs are probabilistic tools, it is not possible explain with complete certainty why a recommendation is made for a particular respondent. As Gemenis highlights, the use of AI in VAAs can cause 'opaqueness', making it difficult to understand how recommendations arise (Gemenis, 2024). Moreover, developers impose values on LLMs either from the top-down, or via reinforcement learning with human feedback (González Barman et al., 2025). The effect that these imposed values have on the LLM's party recommendations is all but impossible to trace, potentially undermining trust in the political advice of an LLM. Related to the matching algorithm, another issue has to do with the LLM's knowledge of party positions. Previous research indicates that off-the-shelf LLMs struggle to position Germany parties, when the benchmark is parties' own responses to the *Wahl-O-Mat* (Batzner et al., 2025). Unfortunately, it is difficult to explain the causes of these inaccuracies. In addition, because off-the-shelf LLMs have a knowledge cutoff date, they are unlikely to capture policy commitments announced by parties during the campaign itself, unless prompted to search online for further information.

*Previous Studies*

Concerns about transparency are somewhat mitigated by developers publishing details about the LLM's training process, or by making visible the LLM's reasoning. However, there will likely always be some degree of uncertainty about the source of LLM output. One way of increasing trust in the outputs of LLMs is test them against some kind of benchmark (e.g. Batzner et al., 2025). Even if the process lacks transparency, we might at least be confident that the LLM can, for instance, mimic with a high degree of reliability the functions of a traditional VAA. More generally, we argue that the best way to answer our research question – *should* LLMs be used as VAAs? – is to simulate how an LLM would handle real data about voter preferences and priorities, and whether a) its advice would lead to a net gain in congruence if followed by voters, and whether b) its advice is similar to that offered by a more traditional matching algorithm (i.e. a salience-weighted proximity model).

Since the release of ChatGPT, a rapidly growing literature has used LLMs to simulate survey responses. Generally speaking, the goal of these studies is to construct 'silicon samples', using data about voters' social identification and political beliefs to anticipate their responses to other questions in the survey (Argyle et al., 2023). While advocates argue that silicon sampling could provide a low-cost way of measuring public opinion, doubts have been raised about LLM performance in contexts that are underrepresented in the training data (Bisbee et al., 2024), as well as a general 'machine bias' (Boelaert et al., 2025). Surprisingly, hardly any studies have used LLMs to predict vote choice itself, with previous work generally using the vote to predict attitudes on other issues (Gudiño et al., 2024). A notable exception is a recent article by von der Heyde et al, who used voters' social identification, left-right position, and party identification to predict the vote in the 2017 German federal election (Von Der Heyde et al., 2025). In predicting the vote, von der Heyde et al finds that ChatGPT 3.5 gives a lot of weight to party identification, while overestimating support for the German Greens and Die Linke (Von Der Heyde et al., 2025). The key difference between our inquiry and that of von der Heyde et al, is that rather than trying to predict the vote, we are asking the LLM which party the respondent *should* vote for – essentially, treating the LLM as a VAA. As a result, we only provide the LLM with the voter's policy preferences and priorities, rather than information about their social or party identification. Moreover, our analysis differs because we benchmark the LLM's outputs not only against the actual party chosen by the respondent, but crucially, against the recommendation made by a salience-weighted proximity model derived from expert perceptions of party positions. This way, our analysis is geared towards testing whether LLMs should be used as VAAs, rather than just asking whether LLMs can successfully predict the next token.

## Research design

### *Data*

The analysis draws on data from the German Longitudinal Election Study (GLES) survey undertaken after the 2025 German federal election. The key advantage of this survey is the inclusion of three policy issue items – tax-spend, immigration, and climate change – that ask the respondent not only about their preference, but also about the personal importance of the issue (Table 1). In each case, the respondent is asked to state their preference on a 1 to 11 scale, and report how important the issue is to them personally on a 1 to 5 scale. Therefore, we are concerned with who the LLM instructs the voter to support when provided with this information about their preferences.

*Table 1. Position issues included in the GLES.*

| Issue | 1 (position) | 11 (position) |
|---|---|---|
| Tax-spend | "Lower taxes and fewer social services" | "More social services and higher taxes" |
| Immigration | "Facilitate immigration for foreigners" | "Restrict immigration of foreigners" |
| Climate change | "Fight against climate change should take precedence, even if it impairs economic growth" | "Economic growth should take precedence, even if it impairs the fight against climate change" |

In total, 7,337 respondents took the survey. However, the analysis is restricted to those 6,022 respondents who reported voting in the federal election. This is because our goal is to test whether LLMs can improve democratic outcomes – for non-voters, there is no 'outcome' we can compare the LLM's recommendation to. More specifically, we use the party vote ("Zweitstimme") as the true indicator of which party the respondent supports. Put simply, this is because the party vote invites the voter to support a *party* at the national level, whereas the constituency vote might also be influenced by local effects (e.g. candidate quality), which are beyond the scope of the investigation. Lastly, the survey includes more variables such as the respondent's gender, age, social class, education, and political interest. In each case, it is interesting to see whether the usefulness of the LLM's recommendation varies – for example, civic engagement tools like VAAs tend to benefit those who have high levels of political interest already (Van de Pol et al., 2014).

*Measures*

Broadly speaking, the goal of the research design is to test whether LLMs can lead to better democratic outcomes, where 'better' means that voters support the party whose positions on important issues are best aligned with their own. Therefore, we must establish benchmarks against which the quality of the LLM recommendations can be evaluated. To begin, we start with the measure of voter-party *congruence*, which is based on the total importance-weighted issue distance between a voter and a party (Equation 1). Essentially, Equation 1 measures how far a voter is from party *j* on *K* policy issues, where each issue term is weighted by its personal importance to the voter *w*. For example, the congruence when a voter is in perfect agreement with a party on all issues would be zero. The more the voter disagrees with a party's positions, the more negative the score becomes, with disagreement having a greater (negative) effect on more important issues.

$$Congruence_j = -\sum_{i}^{K} w_i \cdot |x_i - z_{ij}|$$

*Equation 1. Voter-party congruence is the distance between the voter's position x and the party's position z on K issues of salience w (e.g. Enelow & Hinich, 1984).*

For each voter, the potential congruence is calculated for each of the seven German parties (CDU, SPD, FDP, Green, Die Linke, AfD, and BSW), where party positions are obtained from the 2024 version of the Chapel-Hill Survey (CHES).[1] Previous work shows that measures of congruence based on expert perceptions are positively correlated with traditional VAAs (Gatti, 2026), and so in the circumstances, are probably the closest we can get to approximating what a traditional VAA would recommend with the same data. Following this, two benchmarks are established. First, we take the difference between the congruence of the party the respondent actually voted for ('Congruence vote'), and the congruence of the party the LLM instructed the respondent to vote for ('Congruence LLM'). This yields a measure of how much congruence the respondent would have gained across the three policy issues if it had consulted the LLM about who to vote for, and followed its recommendation (Equation 2). Therefore, whether the Congruence Gained is positive is a first test of whether LLMs could improve voter-party congruence.[2]

[1] In the CHES, country experts are asked to evaluate party positions on scales which match the three issues – tax-spend trade-off, immigration levels, and the economy-climate change trade-off.

$$Congruence_{GAINED} = Congruence_{LLM} - Congruence_{Vote}$$

*Equation 2. Congruence gained (versus the actual vote) by following the LLM's recommendation.*

Second, we take the difference between the congruence of the party the LLM instructed the respondent to vote for ('Congruence LLM'), and the congruence of the most congruent option ('Congruence Max'). In the ideal situation, the voter has perfect information about party positions,

---

These scales are corrected so that instead of running from 0 to 10 (as in the original CHES data), they run from 1 to 11 to match the GLES responses. Moreover, the tax-spend item in the GLES is reversed so that higher values correspond to preferences for lower taxes / reduced state spending.

[2] Clearly, voters have reasons for choosing parties that go well beyond congruence on the three issues included in our analysis. It is not our goal to constrain the question of representation to these three policy issues – far from it. Rather, our interest is in understanding whether an LLM could, when given some basic information about voter preferences, lead voters to a more congruent outcome than before. Therefore, the Congruence Gained measure is just an initial test – if following the LLM's recommendation leads to a congruence loss across the three policy issues compared to the actual vote, it would bode poorly for the chances of the LLM to improve outcomes across a much wider set of issues.

uses this information to estimate their proximity to each party, and chooses the most congruent option. Therefore, the Congruence Max is the best outcome the voter can achieve, and the 'Congruence Missed' represents how much congruence the LLM's recommendation squandered with its choice (Equation 3). Essentially, the Congruence Missed captures the shortfall between the LLMs recommendation and the ideal situation – therefore, it is a sterner test of the LLM's usefulness. If the Congruence Missed is mostly zero, it is an encouraging sign for the LLM's performance.

$$Congruence_{MISSED} = Congruence_{MAX} - Congruence_{LLM}$$

*Equation 3. Congruence missed (versus the maximum) by following the LLM's recommendation.*

Finally, a couple of related statistics are also used to evaluate the LLM's performance. First, the percentage of respondents for whom the LLM's recommendation would lead to a gain in congruence compared to the actual vote ('% Congruence Gained'). Second, the percentage of respondents for whom the LLM's recommendation would cost congruence compared to the ideal vote (i.e. Congruence Max) – in other words, this is the fraction of respondents for which the LLM's recommendation is the same as that of the salience-weighted proximity model ( '% Congruence Missed'). By looking at the percentage of respondents rather than the average Congruence Gained / missed, we ensure that the analysis is robust to extreme cases.[3] Again, it is the Congruence Missed measure which provides a sterner test of the quality of the LLM's recommendations.

[3] Moreover, the population average of the Congruence Gained measure does not tell us anything about the fraction of voters for whom the LLM's recommendation would have a positive effect. In other words, it matters whether the LLM's recommendation a) improves the outcomes of 10% of respondents by 90%, or b) improves the outcomes of 90% of respondents by 10%. In this hypothetical, scenario b) is surely preferable from a democratic point of view.

*Method*

To obtain the LLM's recommendations, we first reduce the original dataset to only the relevant columns – namely, the respondent's id, their issue preferences, priorities, and the party they voted for. We convert the dataset into TSV format, and divide it evenly into 354 files of 20 respondents each.[4] Then, we upload each of the 354 files sequentially to the LLM – Gemini 3 Flash Preview, along with a prompt that explains the task. More specifically, the LLM is asked to instruct each respondent in the uploaded file which party they *should* vote for in the 2025 German federal election, based on the respondent's issue preferences and priorities (for the full prompt, see the Supplemental Material). Finally, the LLM incorporates the recommendation into the TSV file as an additional column.

Subsequently, the congruence of the respondent's actual vote, the LLM's recommendation, and the theoretical maximum congruence are calculated based on Equation 1. Moreover, sociodemographic variables – which are purposefully omitted from the LLM data gathering phase – are integrated back into the dataset. The analysis of the final dataset takes place in three steps. First, we simply analyse the differences between the respondent's actual vote ('Actual vote'), the LLM's recommendation ('LLM model'), and the vote *if* the respondent voted for the most congruent party according to the CHES data ('Expert model'). Second, we calculate the Congruence Gained and Missed by the LLM based on Equations 2-3, and describe how these outcomes vary according to party support. Finally, we use ordinary least squares regression to estimate how the Congruence Gained and Missed vary by the sociodemographic variables, once party support is controlled for. By doing this, we can judge how effective the LLM's recommendations are for different groups of voters.

[4] We argue that this batched approach strikes a compromise between accuracy and efficiency. Given that there are 7,337 survey respondents in total, prompting the LLM separately for each respondent would be too time consuming. However, prompting the LLM only once would make the task too demanding.

## Results

### *Descriptive analysis*

Figure 1 shows the percentage of the sample according the actual vote, alongside the Expert and LLM vote recommendations. At face value, the closeness of the CDU / CSU, SPD, and Green party percentages is impressive. However, we can also see a couple of stark discrepancies. First, the AfD vote (and to a lesser extent, the FDP) is substantially 'overestimated' by the LLM, with 29% of respondents instructed to vote for the far-right party, compared to just under 15% in the actual sample. Second, the vote for Die Linke (and to a lesser extent, the BSW) are severely 'underestimated' by the LLM, with under 2% of respondents instructed to vote for the far-left party. Clearly, the three issue scales used in our analysis – tax-spend, immigration, and climate change – can only capture a fraction of all the variables that might affect the vote. For instance, we cannot account for the reputational damage that belonging to the 'traffic light' coalition government did to the FDP, nor can we account for the stigma attached to voting for the AfD. Therefore, the direct comparison between the LLM's recommendation and the actual vote, while interesting, might just reflect the research design.

Crucially, a fairer comparison *can* be made between the 'Expert' and 'LLM' vote recommendation. As a reminder, the 'Expert' recommendation is the party the respondent would have chosen based on a simple salience-weighted proximity model, where party positions on the three issues are retrieved from the CHES 2024. This means that both models are designed to consider the same criteria – namely, preference and prioritisation of the three issue items. Even here, the LLM recommends voting for the AfD at a far higher rate than the Expert model (29% vs 19%), and Die Linke at a substantially lower rate (2% vs 5%).

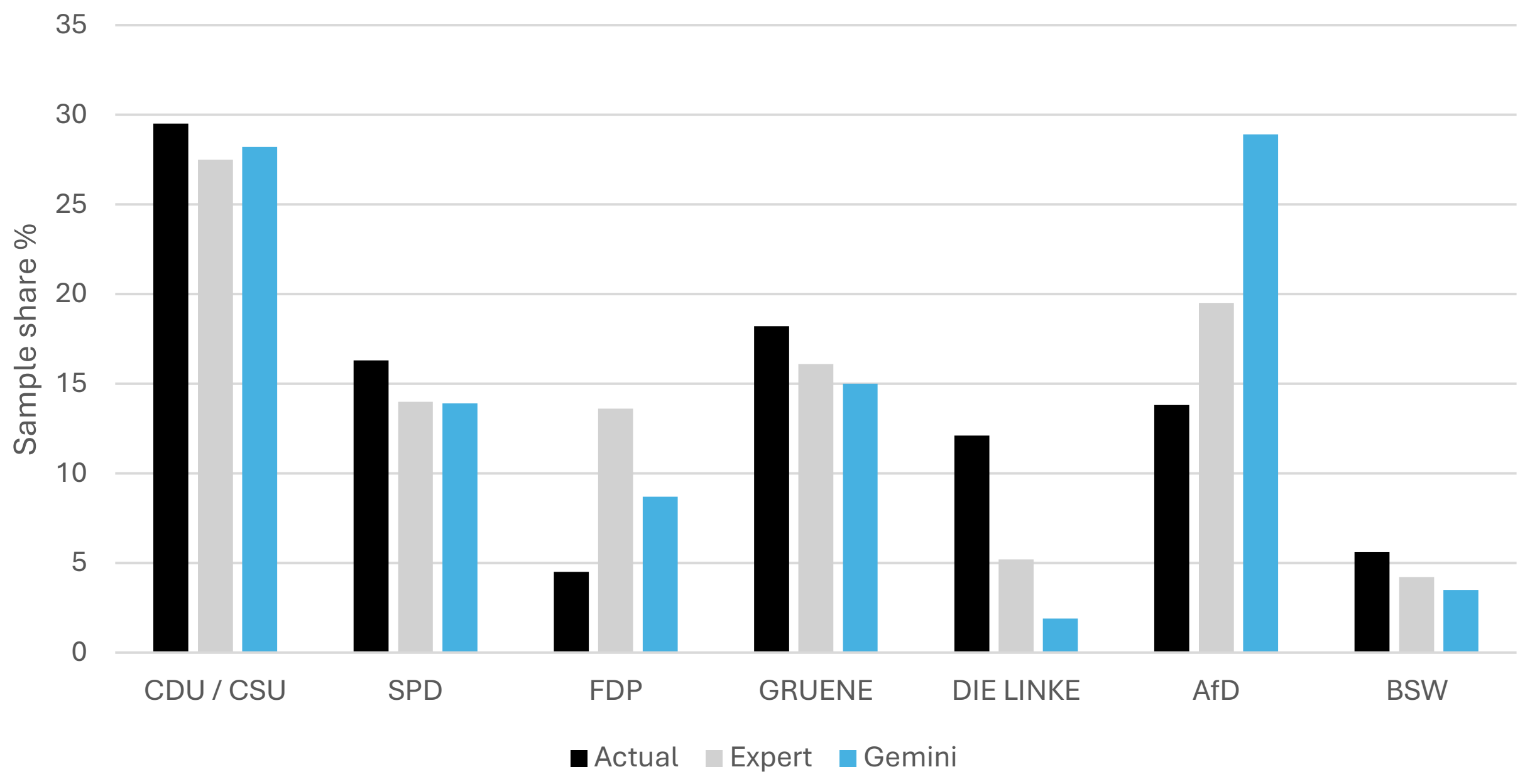

*Figure 1. Share of the actual vote, and the 'Expert' plus 'Gemini recommendations*

The source of these discrepancies is curious, but we can unpack them further by charting the voter flows from the Expert recommendation to the LLM one (Figure 2). Most importantly, we notice that most of the differences occur within ideological 'blocs', where the SPD, Greens, and Die Linke are on the left, while the CDU / CSU, FDP, and AfD are on the right, with the BSW occupying an ambiguous left-right position. For example, the additional AfD recommendations are overwhelmingly for those who would otherwise be advised to support the CDU / CSU or the FDP by the Expert model. While only a small percentage of the total sample, it should also be noted that a sizeable fraction of the anticipated BSW vote goes to the AfD in the LLM model.

Similarly, it is the SPD that absorbs the bulk of the anticipated Die Linke vote. If we consider the Expert model to be the gold standard, then the LLM's 'mistakes' generally involve putting voters into the correct ideological bloc, but recommending the wrong party within that bloc. The one exception to this trend is the significant flow from the SPD to the CDU / CSU.

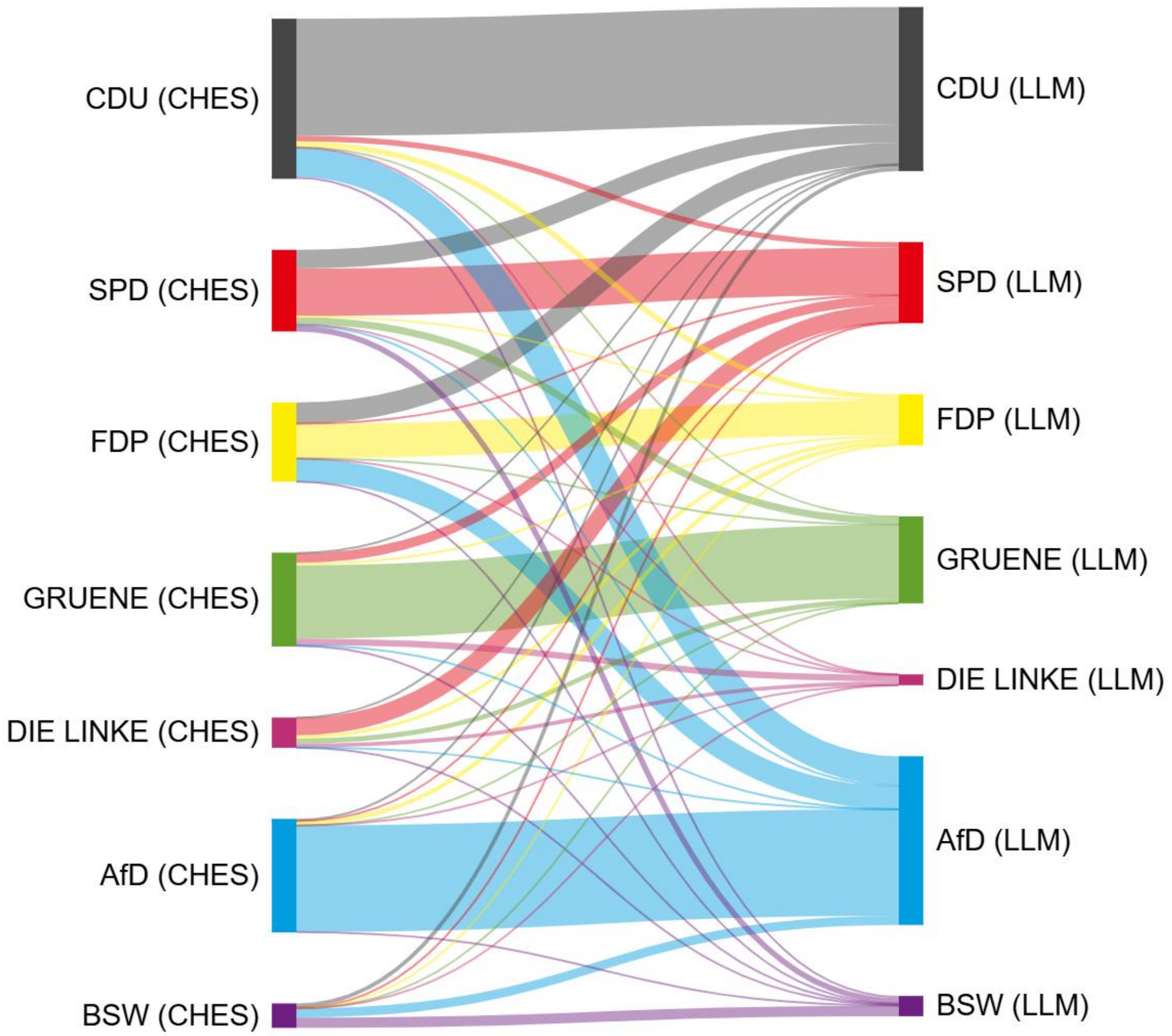

*Figure 2. Mapping the Expert recommendation to the LLM recommendation.*

So far, we have sidestepped the question about the *quality* of the LLM's recommendations. While we have established that they differ substantially from the actual vote (Figure 1), we do not yet know whether these differences would lead to an improvement vis-à-vis congruence. In brief, the analysis shows that the LLM improves the congruence of the majority of voters on the three issues overall ('Congruence Gained'), although still makes an imperfect choice for a significant fraction ('Congruence Missed'). More precisely, the LLM improves the outcome for just over 50% – for these voters, if they had followed the LLM's advice, they would end up being more congruent (for the distribution of Congruence Gained, see Figure 3). For just over a third, the actual vote was the same as the LLM's choice, making the Congruence Gained equal to zero. For just over 12%, the LLM's advice would have led to a net loss in congruence for the voter. On average, the LLM's recommendation results in 7.84 Congruence Gained. Because the Congruence Gained measure combines information about voter preferences and priorities, its values are not so easy to interpret. However, 7.84 is the equivalent of saying that, by following the LLM's advice, the voter would be roughly 1.5 points more congruent with their chosen party (on the 1 to 11 scale) on an issue of maximum importance. To summarise, the results are encouraging to the idea that LLMs could help voters choose parties whose preferences align with their own.

Nonetheless, for 32% of voters, the LLM's recommendation was suboptimal when compared with the maximum potential congruence (for the distribution of Congruence Missed, see Figure 4). In these cases, the LLM's recommendation departs from the Expert model. This could be because the LLM has accessed access to different information about party positions, or because it has introduced other criteria into their 'decision' about which we can only hypothesise.

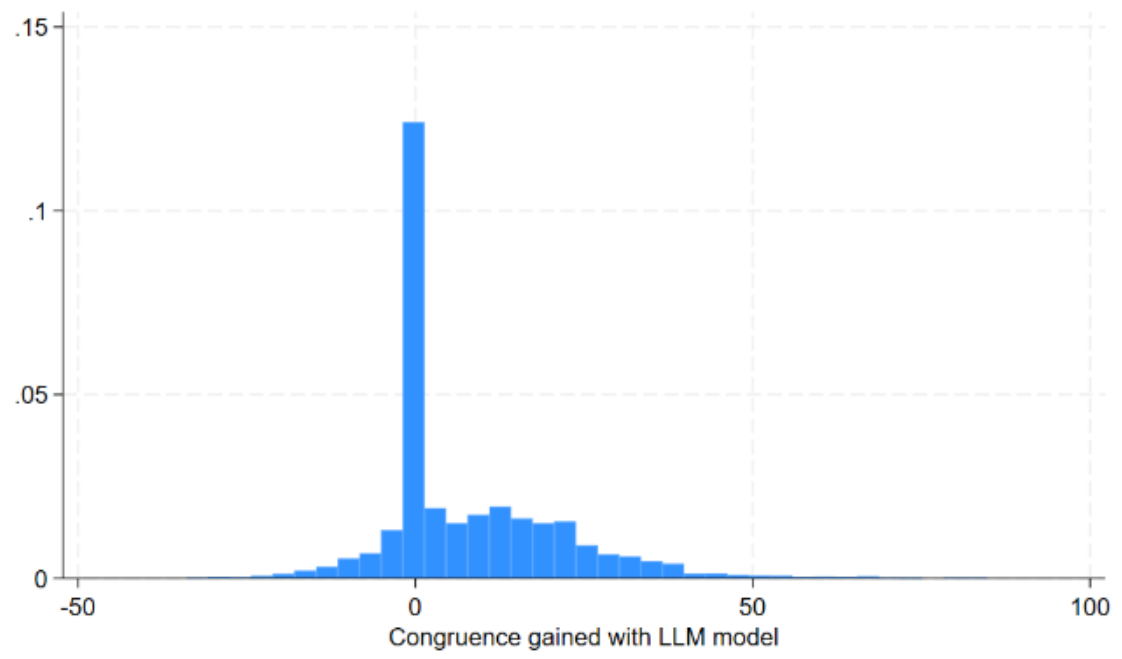

*Figure 3. Distribution of Congruence Gained.*

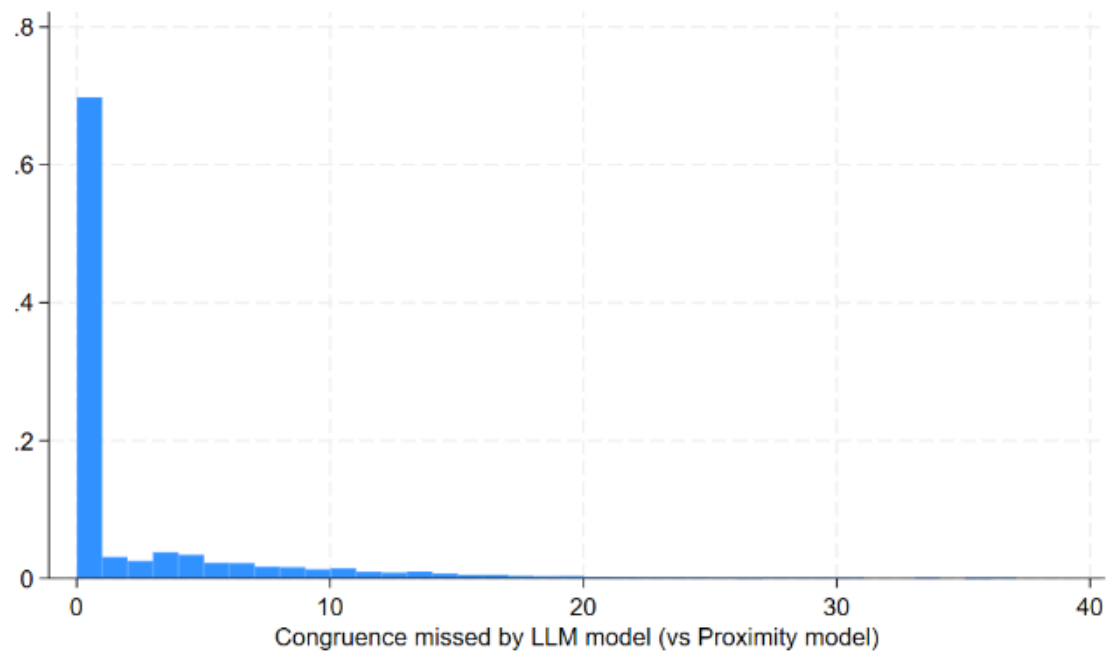

*Figure 4. Distribution of Congruence Missed.*

In the third part of the analysis, we explore how the quality of the LLM's recommendations vary between different voter groups. We begin by looking at how they vary according to the actual vote (Figure 5). We have already seen that for both Die Linke and BSW voters, the LLM is far more likely to recommend a different party (Figure 1). As Figure 5 shows, these recommendations have a net positive effect on congruence on average – however, Die Linke and BSW voters would also squander the most congruence by following the LLM's advice rather than that of the Expert model. On the other hand, the LLM has makes the least difference for AfD voters – in these cases, the agreement between the LLM and Expert models is particularly high, so almost no congruence is missed by the LLM.

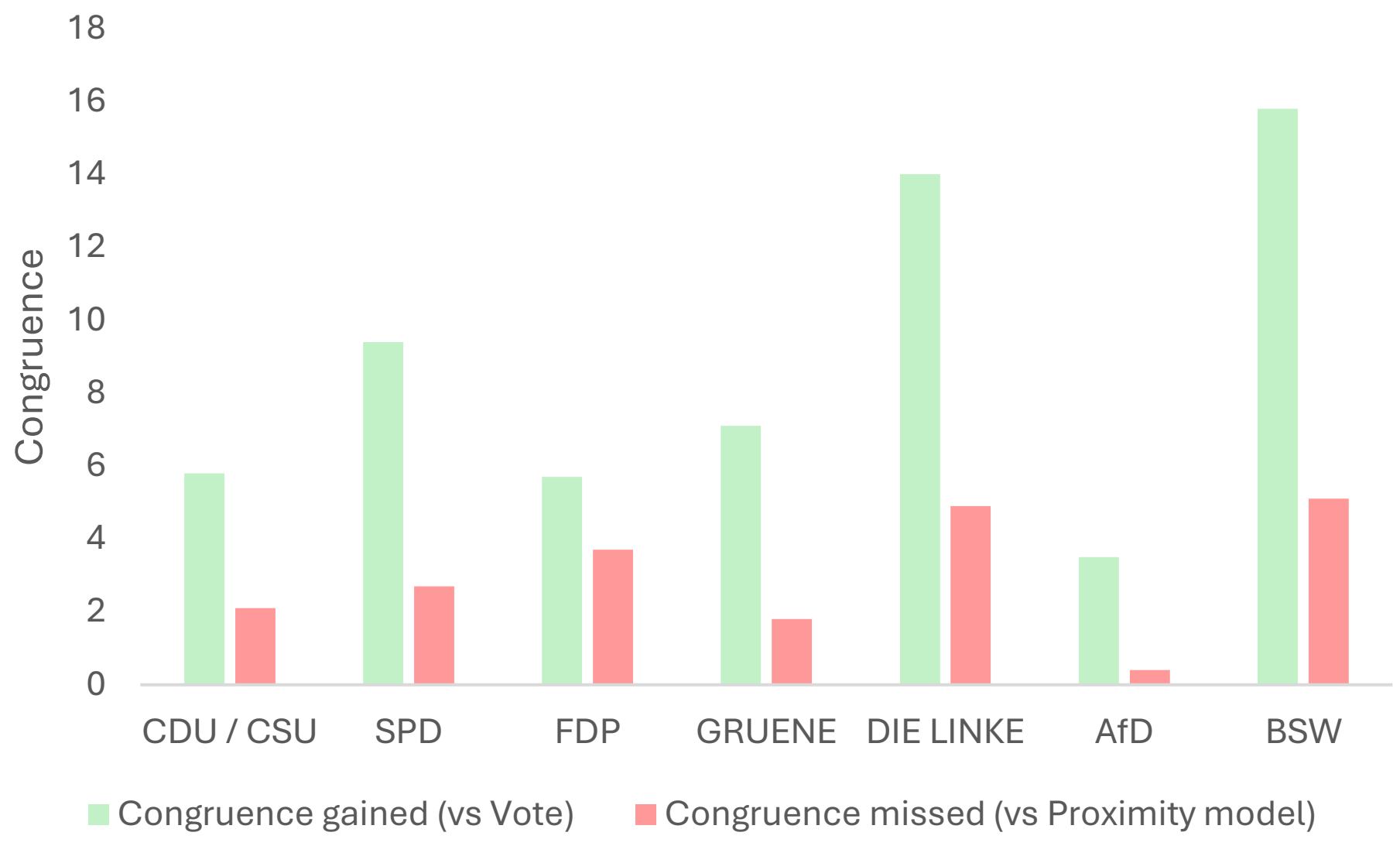

*Figure 5. Congruence Gained and Missed by actual vote.*

Finally, we are also interested in whether the LLM's recommendations would be more useful for certain groups of voters, even once the discrepancies documented in Figure 5 are accounted for. In this regard, the most striking result is that the LLM has the strongest benefits for voters with low levels of political interest and formal education (Table 1). More specifically, the predicted Congruence Gained with the LLM is 43% higher for respondents with the lowest level of political interest compared to the highest (Figure 6). Similarly, the predicted Congruence Gained is 32% higher for respondents with the lowest level of formal education compared to the highest (Figure 7). This is unsurprising, because voters with higher levels of political interest and education should already have good knowledge of party positions, and relatedly, more motivation to vote for parties based on policy agreement (i.e. political sophistication; Luskin, 1987). Therefore, the *remedial* effect of the LLM is stronger among those with low with low levels of political interest and formal education. To summarise the rest of the column, more congruence is gained with the LLM for older respondents, although there is no significant effect for gender nor social class (Table 1).

By contrast, we witness just one significant bias for the Congruence Missed by the LLM compared to the Expert model – namely, the LLM is slightly more likely to give older voters a suboptimal choice (Table 2). Indeed, the R-squared value of this model is very low (0.01), indicating that the variables included in the model do not substantially explain why for some voters, the LLM's recommendation is 'worse' than the Expert one.

*Table 2. OLS regression models of Congruence Gained and Congruence Missed.*

| | Congruence Gained with LLM (vs actual vote) | Congruence Missed by LLM (vs Expert model) |
|---|---|---|
| Gender | -0.599 | -0.0919 |
| | (0.351) | (0.129) |
| Age | 0.523*** | 0.113* |
| | (0.128) | (0.0470) |
| Social class | 0.193 | 0.0779 |
| | (0.186) | (0.0683) |
| Political interest | -0.726*** | 0.0581 |
| | (0.200) | (0.0736) |
| Education | -1.615*** | -0.00294 |
| | (0.280) | (0.103) |
| **Party vote (Reference: CDU / CSU)** | | |
| SPD | 3.502*** | 0.277 |
| | (0.526) | (0.193) |
| FDP | 0.260 | -0.144 |
| | (0.849) | (0.312) |
| GRUENE | 2.568*** | -0.510** |
| | (0.518) | (0.190) |
| DIE LINKE | 9.486*** | 0.0520 |
| | (0.602) | (0.221) |
| AfD | -2.112*** | -0.680** |
| | (0.566) | (0.208) |
| BSW | 10.14*** | 0.240 |
| | (0.798) | (0.293) |
| Constant | 10.02*** | 1.667*** |
| | (1.302) | (0.478) |
| | | |
| Observations | 5,697 | 5,697 |
| R-squared | 0.09 | 0.01 |

Standard errors in parentheses

*** p<0.001, ** p<0.01, * p<0.05

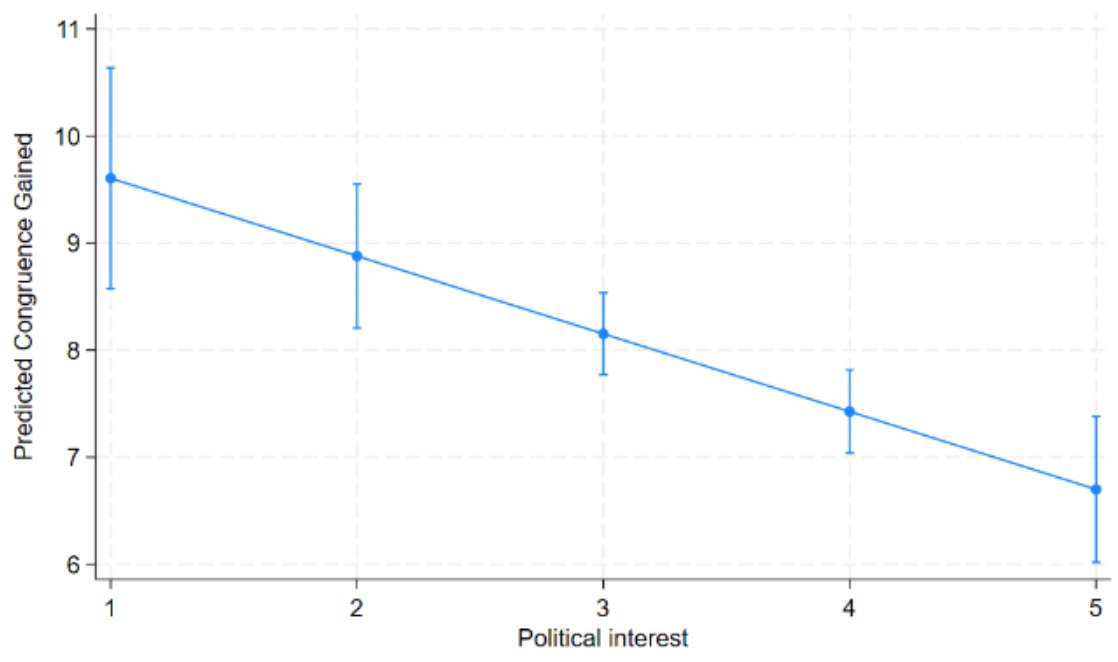

*Figure 6. Predicted Congruence Gained by political interest.*

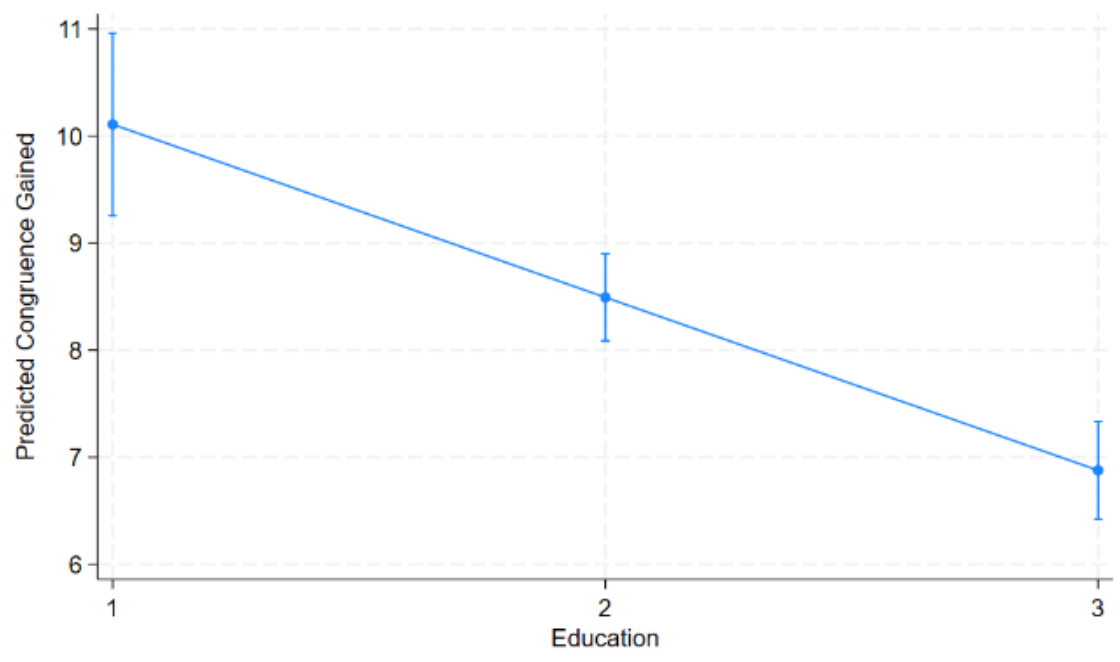

*Figure 7. Predicted Congruence Gained by education.*

## Discussion and conclusions

Can LLMs improve voter-party congruence by assisting voter decision-making? This article addresses this question by prompting Gemini to recommend a party to 6,022 respondents of the most recent GLES, based on their policy preferences and priorities. The analysis resulted in two findings that merit further discussion. On the positive side, for just over half of respondents, the LLM's recommendation would have improved (overall) congruence for the three issues included in the analysis – tax-spend, immigration, and climate change. To be clear, this does not mean that all these voters would 'benefit' had they voted for the LLM's recommendation, as there are dozens of both policy and non-policy criteria that could serve voters' interests which we were not able to include in the model. However, as a proof-of-concept, we show that if given information about a voter's issue preferences and priorities, off-the-shelf LLMs can do a reasonable job of directing the voter towards a congruent party. For just under 70% of voters, the LLM made the same choice as a salience-weighted proximity model where party positions were obtained from the CHES 2024. Therefore, we argue that in making these recommendations, Gemini 3 Flash is broadly capable of replicating a VAA, such as the *Wahl-O-Mat*.

Most impressively, the LLM's advice had a much stronger remedial effect on congruence among those with low levels of political interest and formal education. This is no coincidence, in that it is precisely these voters who are least likely to be congruent with the party they support to begin with (Boonen et al., 2017; Carroll & Kubo, 2018; Enyedi et al., 2020; Lesschaeve, 2017; Rodríguez-Zepeda & Otero-Felipe, 2021). In relation to this, one issue with VAAs is that they tend to attract a more politically interested audience, who in many cases will already be aware of which parties best represent their policy preferences (Van de Pol et al., 2014). However, with freely available

LLMs, the voter need not go to the effort of seeking out a bespoke application – therefore, this approach might reduce the engagement barrier faced by VAAs.

However, on the negative side, the analysis revealed that there are substantial biases in the recommendations made by the LLM. In particular, the LLM was much more likely to recommend voting for the AfD than the Expert model. More precisely, 29% of the sample was instructed to vote for the AfD, compared to just 20% in the Expert model, and only 15% who reported actually voting for the far-right party.  On the contrary, the LLM rarely ever instructed voting for either Die Linke or the BSW. This echoes the finding that LLMs tend to underestimate support for smaller parties (Von Der Heyde et al., 2025). We cannot be certain about the source of these discrepancies – although, there are a couple of plausible explanations. First, because they are a controversial party whose rise has received extensive media and scholarly attention, the AfD might simply be better represented in the LLM's training data (e.g. Kandpal et al., 2023). By contrast, Die Linke and the BSW are smaller parties whose names probably appear less frequently. Second, the LLM might have different perceptions of party positions compared to the CHES – hypothetically, evaluating the CDU / CSU as more moderate on the immigration theme than the country experts would nudge more anti-immigration voters in the direction of the AfD. In more general terms, the LLM could be making considerations that go beyond a salience-weighted spatial logic – for instance, using a directional or discounting logic of voter behaviour (Adams et al., 2005; Rabinowitz & Macdonald, 1989), or introducing its own issue salience scores based on its training data. Another possibility is that GLES respondents were reluctant to admit voting for the AfD, leading to an artificially low number of reported AfD votes. However, this does not explain the difference between the recommendations of the LLM model and the Expert model.

To summarise, there are several reasons why the LLM's advice might diverge from the Expert model, let alone the actual vote. In any case, the lack of *transparency* about how the recommendations are produced undermines the normative value of the tool. If the LLM's recommendations reflect inequalities in the training data, or values imposed top-down by the architects, then the democratic benefit of LLM-assisted voting is dubious to say the least. At worst, LLMs could become tools for manipulation, nudging voters towards the 'right' parties under the guise of neutrality. While the same questions apply to VAAs (and indeed, the Expert model used in this article), at least the methodology of these is usually made transparent, which means that the assumptions underlying them can be criticised.

Lastly, there are a few limitations to the research design for future work to address. First, we would be interested to see how the quality of the LLM's recommendations are affected by having more policy and non-policy information about the respondent. Second, and in close relation to the previous point, LLM-assisted voting will not be as clear-cut in reality as it is in our research design. Users are likely to ask for voting advice without opening a new account, meaning that the LLM's responses will also be biased by the previous conversations. Similarly, the user is unlikely quantify their preferences and priorities in the way we have done. Nonetheless, LLMs tend to 'think' about voting much in the way that VAAs are programmed to – namely, in terms of policy preferences and priorities. Third, while we have decided to focus on Gemini, future work could try to replicate the quantitative analysis in ChatGPT, Claude, Grok, or the other publicly available LLMs. If the findings are not replicated, this would make it even more difficult to correct for the biases we have observed in this article.

**Supplemental Material**

**Part A) Gemini prompt**

*Imagine that each row in uploaded table corresponds to a respondent ("voter_unique_ID") who asks you for guidance about who to vote for in the 2025 German federal election.*

*Note - German voters vote twice in federal elections. The first vote is for a candidate in their local constituency, the second vote is for a party. I want you to advise each respondent about who should receive their party vote ("Zweitstimme").*

*The table includes each respondent's preferences about three political issues - "taxes_vs_social services", "immigration", and "climate change_vs_economic growth". In addition, the each respondent is asked how important they find each issue on a scale of 1 to 5. The exact wording of the survey questions is as follows:*

*Issue = taxes_vs_social_services*

*"Some people prefer lower taxes, although this results in less social services. Others prefer more social services, although this results in raising taxes. What position do you take on taxes and social services?"*

*Scale:*

*(1) 1 More social services and higher taxes*

*(2) 2*

*(3) 3*

*(4) 4*

*(5) 5*

*(6) 6*

*(7) 7*

*(8) 8*

*(9) 9*

*(10) 10*

*(11) 11 Lower taxes and fewer social services*

*How important is the topic of taxes and social services to you?*

*Scale:*

*(1) Not important at all*

*(2) Not very important*

*(3) In between*

*(4) Somewhat important*

*(5) Very important*

*Issue = immigration*

*"Now, let's talk about immigration opportunities for foreigners. Some want to facilitate immigration for foreigners, others want to make immigration more difficult for foreigners. What position do you take on immigration for foreigners?"*

*Scale:*

*(1) 1 Facilitate immigration for foreigners*

*(2) 2*

*(3) 3*

*(4) 4*

*(5) 5*

*(6) 6*

*(7) 7*

*(8) 8*

*(9) 9*

*(10) 10*

*(11) 11 Restrict immigration of foreigners*

*How important is the topic of immigration of foreigners to you?*

*Scale:*

*(1) Not important at all*

*(2) Not very important*

*(3) In between*

*(4) Somewhat important*

*(5) Very important*

*Issue = climate_change_vs_economic_growth*

*"And now onto the topic of climate change and economic growth. Some say that the fight against climate change should definitely take precedence, even if it impairs economic growth. Others say that the economic growth should definitely take precedence, even if it impairs the fight against climate change. What is your position on fighting climate change and supporting economic growth?"*

*Scale:*

*(1) 1 Fight against climate change should take precedence, even if it impairs economic growth*

*(2) 2*

*(3) 3*

*(4) 4*

*(5) 5*

*(6) 6*

*(7) 7*

*(8) 8*

*(9) 9*

*(10) 10*

*(11) 11 Economic growth should take precedence, even if it impairs the fight against climate change*

*"How important is the issue of fighting climate change and supporting economic growth to you?"*

*Scale:*

*(1) Not important at all*

*(2) Not very important*

*(3) In between*

*(4) Somewhat important*

*(5) Very important*

*Task = Please complete the last column of the uploaded table "best_match_political_party" by writing the name of the political party the voter should vote for in the 2025 German federal election with their party vote ("Zweitstimme").*

*Your answer must contain the full TSV table, and only that. The TSV table must be perfectly preserved, except for the changes that were explicitly requested. Write only one political party name, and nothing else. Do not write any meta-comment.*

**Part B) Mapping the Actual vote to the LLM recommendation**

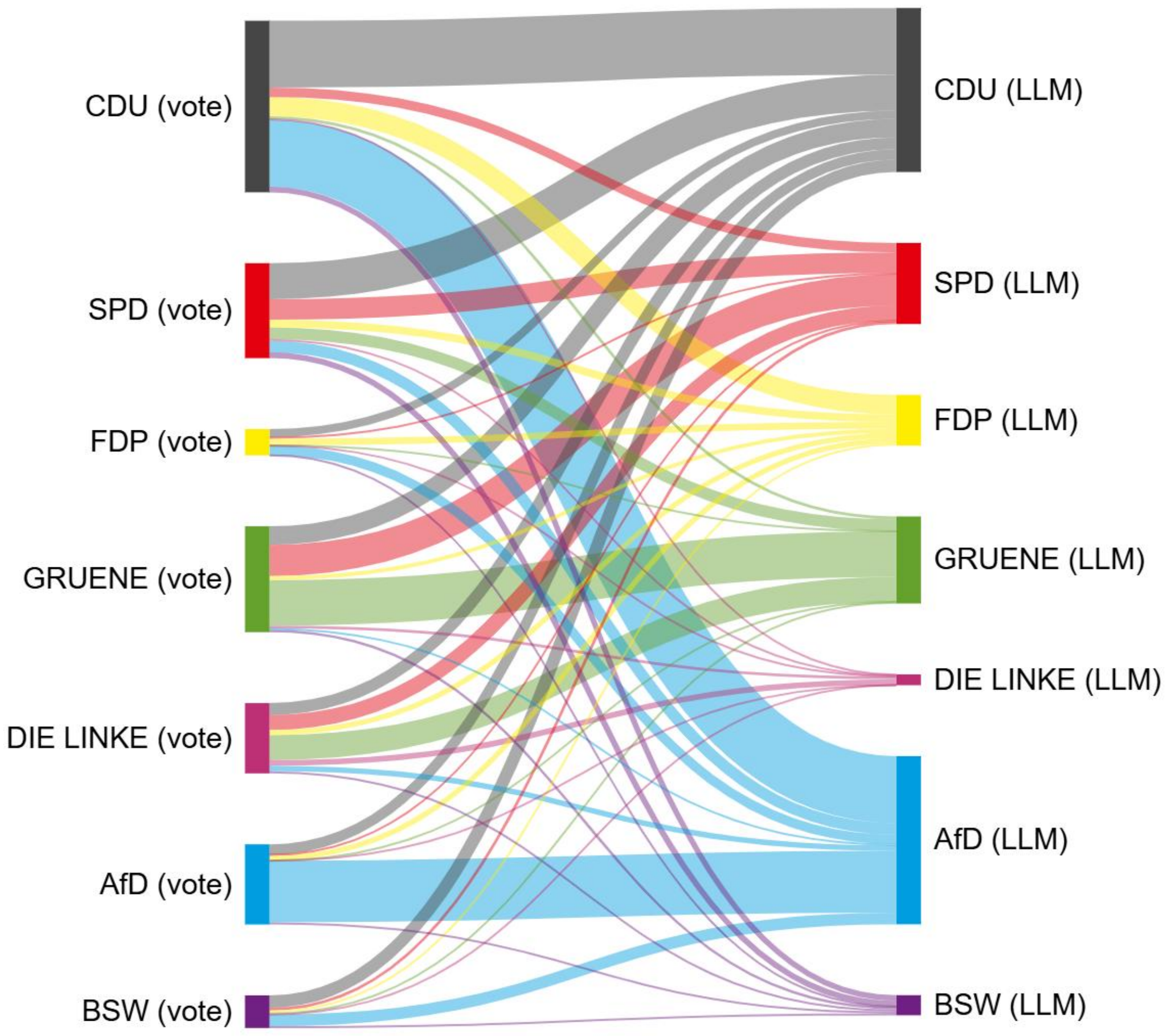

*Figure 2. Mapping the Actual vote to the LLM recommendation.*

.

**Part C) Replication statistics**

Because LLMs are probabilistic, they will sometimes give different answers to the same question. Therefore, one risk in the research design is that the recommendations made by the LLM are mostly due to chance, rather than systematic patterns. We mitigate against this risk by prompting the LLM to give a recommendation to all 6,022 respondents for a second time. In 85% of cases, the party recommended to the responded was the same as in the first attempt. However, at the aggregate-level, the distribution of the responses by party was practically identical (Figure C1). This suggests that the previously discussed party biases are a feature of the LLM's output, rather than being due to chance. Likewise, on the second attempt, the LLM's recommendations were near-identical in the extent to which they agreed with the Expert model (67% of respondents). To summarise, we find no cause for concern with regards to reliability of the findings.

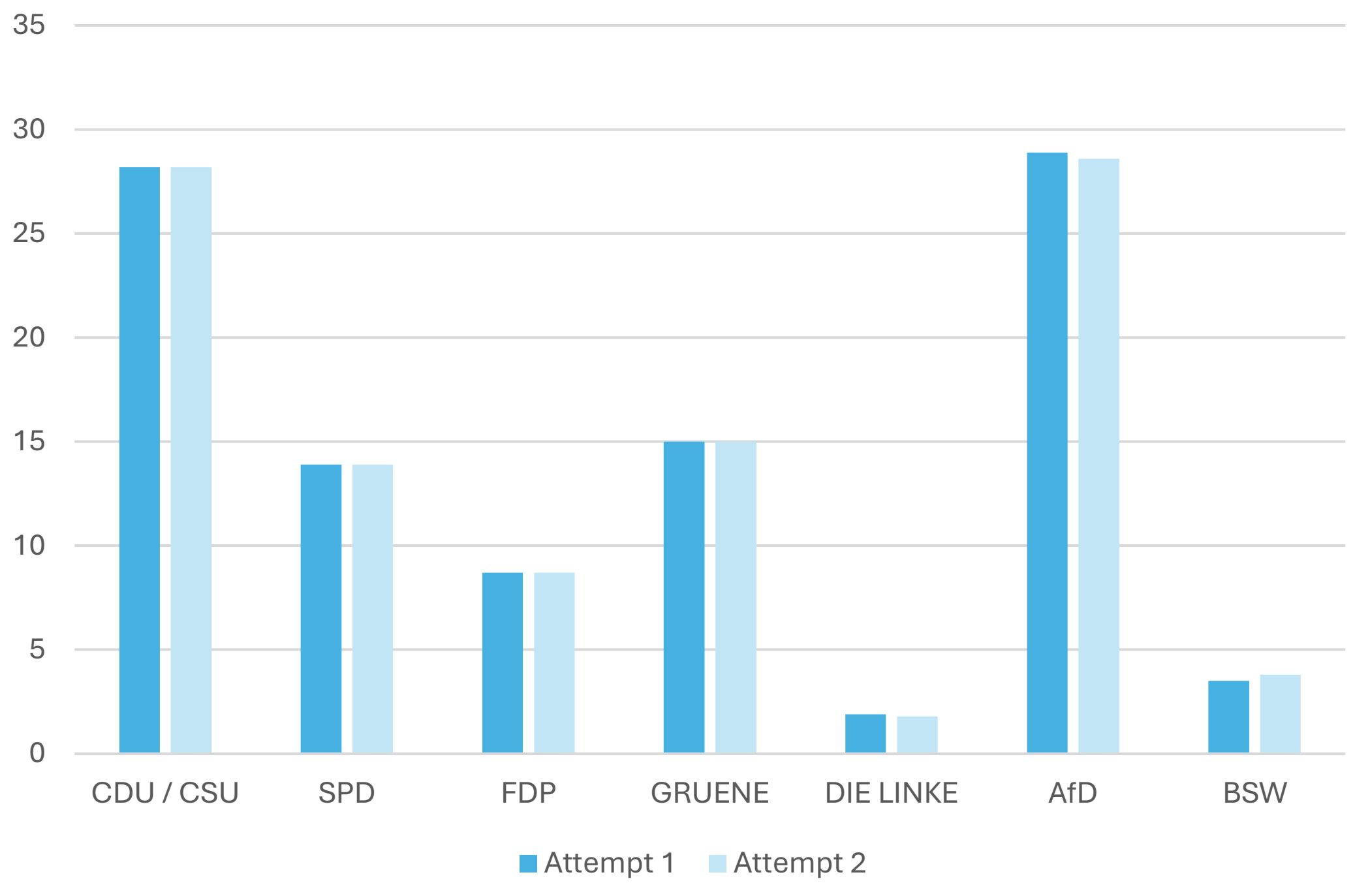

*Figure C1. LLM recommendations on first versus second attempt by party.*